\documentclass[preprint]{ecai} 

\usepackage{latexsym}
\usepackage{amsthm}
\usepackage{color}
\usepackage{times}
\usepackage{graphicx}
\usepackage{amsmath}
\usepackage{amssymb}
\usepackage{hyperref}
\usepackage{booktabs}
\usepackage{enumitem}
\usepackage{algorithm}
\usepackage{algpseudocode}
\usepackage{xcolor}
\usepackage{tikz}
\usetikzlibrary{shapes, arrows, positioning, fit, backgrounds}

\newcommand{\BibTeX}{B\kern-.05em{\sc i\kern-.025em b}\kern-.08em\TeX}

\begin{document}


\begin{frontmatter}


\paperid{4713} 


\title{HECATE: An ECS-based Framework for Teaching and Developing Multi-Agent Systems}



\author{Arthur Casals}
\author{Anarosa Alves Franco Brand{\~a}o}

\address{Universidade de S{\~a}o Paulo, S{\~a}o Paulo, Brazil}


\begin{abstract}
This paper introduces HECATE, a novel framework based on the Entity-Component-System (ECS) architectural pattern that bridges the gap between distributed systems engineering and MAS development. HECATE is built using the Entity-Component-System architectural pattern, leveraging data-oriented design to implement multiagent systems. This approach involves engineering multiagent systems (MAS) from a distributed systems (DS) perspective, integrating agent concepts directly into the DS domain. This approach simplifies MAS development by (i) reducing the need for specialized agent knowledge and (ii) leveraging familiar DS patterns and standards to minimize the agent-specific knowledge required for engineering MAS. We present the framework's architecture, core components, and implementation approach, demonstrating how it supports different agent models.
\end{abstract}

\end{frontmatter}


\section{Introduction}
In 1995, Russell and Norvig~\cite{Russell:1995:AIM:193191} stated that ``AI is the study of agents". An agent is a computer system with autonomous capabilities, able to make decisions on what needs to be done to satisfy their goals. Agents can be organized in communities known as \textit{multiagent systems} (MASs): systems composed of multiple agents that interact among themselves in a single environment~\cite{Russell:1995:AIM:193191} to solve complex problems. Regardless of their nature, problems in MAS may have different complexity degrees, and they can be divided into sub-problems (or goals) individually assigned to each agent in the system~\cite{ferber1999multi}.

Due to their inherently distributed mechanisms, it is easy to trace a parallel between MAS and \textit{distributed systems}. A distributed system (DS) is ``a collection of autonomous computing elements that appears to its users as a single coherent system"~\cite{steen2017distributed}. In practice, DSs are systems in which components located on networked computers communicate and coordinate their actions by passing messages. Similarly to MAS, their components concurrently interact with one another to achieve a common goal. Also, their computing elements (\textit{nodes}) can behave independently. The evolution of both research fields from Networked Computing was always close, with influences and convergences over time~\cite{chopra2021multiagent}. 

Multi-agent systems (MAS) have evolved alongside distributed systems (DS) for decades, sharing common foundations in decomposable, concurrent, and autonomous computing. However, while distributed systems engineering has been widely adopted through architectural patterns and frameworks in industry, MAS development remains largely confined to academic settings and specialized applications~\cite{Mascardi2019}. While we have well-established MAS development platforms and frameworks in the agent community, such as JaCaMo~\cite{boissier2013multi}, JADE~\cite{bellifemine2007developing}, and SPADE~\cite{palanca2020spade}, their use is also almost exclusive to the agent community. Most of them have not been used yet to deploy a MAS into production~\cite{mascardifantastic}, being used primarily in academic research. In addition, none of the most popular MAS frameworks eliminate the need for specialized knowledge such as goals or beliefs (in the case of BDI~\cite{rao1995bdi} agents) or how the system can be implemented considering different organizations, artifacts, and collaborating agents. Finally, there are practically no engineering standards for MAS (except for communications) - a huge difference when compared with distributed systems~\cite{steen2017distributed}.

In this context, integrating MAS concepts with established DS architectures could potentially provide a powerful combination to harness the benefits of both paradigms when implementing MASs. The decentralized nature of DS supports the distributed intelligence of MAS, facilitating seamless communication and coordination among agents. Taking advantage of the already-established standards and tools used by DS could greatly minimize the complexity of specialized MAS frameworks. Furthermore, incorporating the latest AI advancements into newly developed MASs could make it much easier, thus significantly enhancing their decision-making, learning, and adaptive capabilities.

Our research motivation lies in bridging the gap between MAS and DS, leveraging the strengths of both fields. Our ultimate goal is to develop practical methods and tools for engineering sophisticated MAS to meet the demands of modern software systems. We trace a parallel between DS and MAS in our research because while they share multiple elements in their history, the software engineering community sees and uses both differently. 


In this paper, we present HECATE, a framework that implements this approach by mapping multi-agent system concepts to the Entity-Component-System (ECS) architectural pattern. HECATE addresses two key challenges: (i) leveraging existing distributed systems techniques and tools for MAS development, and (ii) reducing the learning curve for students and practitioners new to agent systems. 




This paper is organized as follows: Section \ref{sec:background} provides essential background on MAS concepts and ECS architecture, as well as existing work related to our research. Section \ref{sec:design} presents the framework requirements and design. Section \ref{sec:architecture} details the HECATE architecture. Section \ref{sec:implementation-overview} describes the implementation approach. Section \ref{sec:case-study} briefly discusses the case studies involving HECATE. Finally, in Section \ref{sec:discussion} we discuss the potential benefits of our approach, outlining the main conclusions and future work.

\section{Background and Related Work} 
\label{sec:background}

\subsection{Engineering multiagent systems}
Software engineering abstractions were explored in the domain of agents. Recently, the authors in~\cite{chopra2022interaction} proposed a set of programming abstractions for autonomy and decentralization of MAS - also touching the subjects of microservices, communication, and programming models for distributed systems. A more encompassing approach - and also closer to our research - was proposed by the authors in~\cite{silva2002taming}, with a conceptual framework named TAO built upon object and agent-based abstractions. Such a conceptual framework was the basis for the minimal set of abstractions to our proposal.
 

In 2007, a survey on existing agent methodologies based on the agent-oriented paradigm was published~\cite{cabri2007service} with the aim of evaluating to what extent existing agent-oriented methodologies were also oriented to the development of services. From the Agent-oriented Software Engineering (AOSE) perspective, the study demonstrated that all evaluated AOSE methodologies could be used to model service-oriented agents. At this point, the focus was on modeling agents as services or generally using the concepts around services to facilitate the interaction between agents. 

Since our research topic is closely related to the existing MAS tools and platforms to support engineering MAS, 
we frequently cite three of the most popular MAS development platforms and frameworks: JaCaMo~\cite{boissier2013multi}, JADE~\cite{su2011jade}, and SPADE~\cite{palanca2020spade}. Their relevance is established in different publications~\cite{mascardifantastic,leon2015review}, and they are the best examples of incorporating DS techniques into the agent domain.

\subsection{Multiagent systems and Distributed systems architectures}
From the architectural perspective, the common roots between MAS and DS have also been discussed extensively. They are clear when we focus on the subject of MAS architectures~\cite{weiss2016multiagent,khosla2013design,hopgood2021intelligent,weyns2010architecture}. In~\cite{chopra2021multiagent}, Chopra \textit{et al.} touch the common evolution of both domains, introducing a series of architectural abstractions meant to show how multiagent systems can be used to address problems from the distributed systems' domain. While this is a recent position paper, the architectural parallel between MAS and DS has also been explored from different perspectives long before, especially in communications~\cite{ahmad2002multi,dzitac2009artificial,nissim2012multi,lange1998mobile,christie2023kiko,baldoni2021reimagining}.

The agent community has also addressed modern decentralized DS architectures and their benefits. In~\cite{chopra2021deserv}, the authors propose a serverless agent programming model~\cite{mcgrath2017serverless} for cloud-based MAS. Fault tolerance for decentralized MAS applications is also explored in~\cite{christie2022mandrake}. Other modern architectural topics such as microservices~\cite{newman2015building,khadse2023protocol}, containers~\cite{burns2016design,pfeifer2021multi}, Internet of Things (IoT)~\cite{rose2015internet,fortino2014integration,ciortea2017beyond} have also been explored by the agent community.

Some of these topics also reached the established MAS development tools. In ~\cite{Ciortea:2017:GAR:3091125.3091342}, the authors explore using web communication standards with JaCaMo. They proposed that the Web, in its current state, could be suitable as a middleware for Internet-scale multiagent systems. This would be achieved through the use of a resource-oriented layer that would allow the application environment to be decoupled from its deployment context. To demonstrate this proposal, an agent environment was developed using JaCaMo. This environment was specifically aimed at the Internet of Things (IoT), emphasizing the separation between the application environment and the deployment context.

The platform is also used in~\cite{o2020delivering}, where the authors discuss deploying agents within microservices with their virtual resources implemented as CArtAgO artifacts. This idea, previously explored in~\cite{w2019mams}, is based on the premise that microservices were one of the state-of-the-art architectural styles for large-scale distributed software development~\cite{salah2016evolution}. This premise still holds, and communication between agents is largely addressed in this study. The agents were implemented in ASTRA~\cite{collier2015reflecting}, an agent-oriented programming language closely integrated with Java.

While modern decentralized DS architectures and their benefits have been addressed to some extent by the community, they were not deeply explored. Also, the existing work in this field focuses mainly on bringing or reusing DS architectural aspects in the MAS domain. In our approach, we intend to take the opposite direction: bringing MAS concepts into the DS domain.

\subsection{The Entity-Component-System (ECS) Pattern}

The Entity-Component-System (ECS) pattern is an architectural pattern based on the concept of Component-Based Systems~\cite{crnkovic2002building}, using a \textit{data-oriented design} (DOD)~\cite{fabian2020data}. This is one of the DS architectural patterns most used in complex, massive multiplayer online games (MMOGs)~\cite{nystrom2014game,gregory2017game}.  

The principles behind DOD have been around for a while, but it only received this name in 2009~\footnote{https://gamesfromwithin.com/data-oriented-design}. DOD is a program optimization approach focused on the efficient usage of the CPU cache for manipulating data. As such, it emphasizes data layout, separating it from the problem domain (data is separated from logic). Object-oriented design (OOD)~\cite{meyer1997object}, in contrast, focuses on encapsulating data and attributes within objects. As a result, the data is inherently contextualized (making it part of the problem domain).

There are different studies on the advantages of adopting DOD instead of OOD in high-performance systems~\cite{mironov2021comparison,fedoseev2020case}. In these studies, one of the common conclusions is that using DOD makes it easier for scaling systems that process large quantities of data. This is particularly interesting to MAS development, since in a multiagent system there can be, hypothetically, myriad of agents simultaneously performing actions based on the same data (environment). DOD is frequently used in game development, which is also interesting from the MAS perspective. Both in MASs and games, the system has to be designed to deal with multiple autonomous entities (agents and game characters, respectively). In game development, non-playing characters (NPCs) frequently interact with each other and the environment. Their actions can depend on the environment's state, and can also affect it as a result (most of the times, in real time). There are many other aspects from the agents field that can also be present in game development (such as behavioral models, coordination, cooperation, communication). As a result, when looking for a DS architectural pattern to test our hypothesis, it makes sense to look into the game development field.

The ECS pattern appeared and evolved within the MMOG community~\footnote{http://t-machine.org/index.php/2007/09/03/entity-systems-are-the-future-of-mmog-development-part-1/}~\footnote{https://unity.com/dots} and was quickly adopted by modern Real-Time Interactive Systems (RIS) frameworks~\cite{wiebusch2015decoupling} and commercial game engines~\cite{baron2019hands}. As a DOD-focused pattern, it emphasizes composition over inheritance by decoupling data from logic. It is especially designed for scenarios where many different, independent entities interact dynamically with the environment and each other. 

The ECS pattern is based on three core abstractions:
\begin{itemize}
    \item[$\bullet$] \textit{Entities}: Unique identifiers representing objects in the system. Entities are lightweight containers without inherent behavior or data, serving as aggregators for \textit{components} (below). In a game (or in a MAS), an entity could represent an agent, or a specific element in the environment.
    \item[$\bullet$] \textit{Components}: Pure data containers that define entity attributes. Components are organized for optimal memory access patterns, enabling efficient processing through data-oriented design principles. Each {\tt Component} stores a single type of data, and - as its name implies - they are used solely as data containers, holding no behavior. A single Entity may have multiple {\tt Components} attached to it in order to define its characteristics. In an ECS implementation, {\tt Components} are stored in arrays or contiguous blocks of memory, as a way of promoting cache usage efficiency. Since {\tt Components} are responsible only for storing data, any modifications to data structures or its components do not affect the overall logic of the program.
    \item[$\bullet$] \textit{Systems}: Logic processors that operate on entities possessing specific component combinations. {\tt Systems} implement all behavior, maintaining clear separation between data and logic. They are designed to operate on Entities possessing a specific {\tt Component} composition: within a game, for example, a {\tt System} for processing "movement" can be used to update all {\tt Entities} that possess the combination of {\tt Components} related to "position" and "velocity." 
\end{itemize}

This structure is heavily aligned with DOD principles, emphasizing data organization for optimal processing efficiency. ECS implementations also use the concepts of (i) ``World" (the environment in which the game happens) and (ii) Events to aid the logic behind systems: all entities are susceptible to different events triggered within the world. A Component is extensible, and it modifies each Entity by providing specific data in the form of properties. Considering a game environment, a given Entity may possess a Component called "Position", with properties (X, Y, and Z coordinates) that determine where in the world the Entity is located. While the three core abstractions explained above form the foundation of ECS, additional abstractions and utilities are usually required to allow or enhance the system's functionality, manageability, and performance.  


One crucial aspect of the ECS architecture is the event-based communication used within the system. Events in an ECS architecture serve as the backbone for communication between different parts of the system, ensuring that Systems and entities can interact without being tightly coupled. This decoupling is crucial for maintainability and flexibility.

Events are typically triggered when something significant occurs in the system, such as an entity colliding with another or an entity entering a specific zone. For example, consider a scenario where a physics system detects a collision. Instead of directly notifying all systems that might care about collisions, the physics system can emit a "collision" event. Other management Systems responsible for reacting to the collision can then listen to this event and react appropriately.

Events can be processed in two primary ways:

\begin{itemize}
    \item Immediate Events: These are handled as soon as they are emitted. For instance, when a player picks up a health pack, the health system might immediately increase the player’s health.
    \item Queued Events: These are added to a queue and processed at a later time, often in the next update cycle. This approach is useful for complex systems where actions might need to be deferred to maintain consistency.
\end{itemize}

Managers in an ECS architecture are specialized components that handle the lifecycle and interactions of the ECS's core abstractions—entities, components, systems, and events—ensuring that the system runs smoothly and efficiently.

The \textit{entity manager} is responsible for maintaining a registry of all entities in the system. It assigns unique identifiers to entities and handles their creation and destruction. For instance, let us assume the context of a MAS built using the ECS architecture. Whenever a new agent is deployed into the environment, the entity manager creates a new entity (to identify this agent) and registers it in the system. The entity manager also provides querying capabilities, enabling other systems to find entities based on their components. 

Components store the data that define an entity's characteristics. The \textit{component manager} handles the storage, retrieval, and organization of these components. To optimize performance, components are often stored in contiguous memory arrays to facilitate cache-friendly access patterns. For example, if 1,000 entities have a "position" component, the component manager ensures that all these components are stored together in memory, reducing the overhead of scattered memory access.

The \textit{System manager} organizes and coordinates the execution of Systems. It ensures that Systems are updated in the correct sequence, respecting dependencies between Systems. For instance, in a game, the physics System must update before the rendering System to ensure that visual representations reflect the latest physics calculations. The System manager may also handle system-level optimizations, such as parallelizing independent Systems for multithreaded execution.

Finally, the \textit{event manager} is responsible for managing the lifecycle of events. It maintains a registry of systems or entities interested in specific event types and ensures that events are dispatched efficiently and received by the appropriate listeners.

The ECS engine orchestrates the entire architecture, integrating all the other components to ensure smooth operation. The core loop drives the ECS's functionality and is responsible for initializing the system, managing updates, and shutting it down cleanly.

The engine handles the main update loop, sequentially updating all Systems in the correct order. For instance, during each frame in a game, the engine might:

\begin{itemize}
    \item Update the input System to process player inputs.
    \item Update the AI System to compute NPC behaviors.
    \item Update the physics System to simulate movements and collisions.
    \item Update the rendering System to display the results on the screen.
\end{itemize}

The ECS engine also acts as the primary interface for developers, providing functions to create entities, add or remove components, and manage systems. It encapsulates the ECS's complexity, allowing developers to focus on implementing the desired functionality.

Other common components in an ECS-based system are the \textit{memory manager} and the \textit{logger}. Efficient memory management is critical in an ECS architecture, especially for large-scale systems such as games or simulations where thousands of entities and components must be managed simultaneously. The memory manager optimizes memory allocation and usage to avoid bottlenecks. This can be done using memory pools or arenas, where a fixed amount of memory is pre-allocated for each component type. This reduces the overhead of dynamic memory allocation during runtime, which can be slow and prone to fragmentation. For example, a memory pool for "position" components might allocate space for 10,000 components upfront, ensuring that adding new entities with a position component is instantaneous.

Using a logger is common practice in different kinds of systems. It provides visibility into the system's state and behavior. It is especially useful during development and debugging, where it is crucial to understand the interactions between entities, components, and systems.

The logger records various types of information, such as:

\begin{itemize}
    \item Entity lifecycle events, including creation, updates, and destruction.
    \item Component changes, such as when a health component's value is modified.
    \item System execution, including the order of execution and time taken for each system.
\end{itemize}

For example, if an entity unexpectedly disappears from the game, the logger can reveal whether it was destroyed intentionally or due to a bug. Similarly, if the game experiences performance issues, the logger can pinpoint which system is causing the bottleneck.

Advanced loggers can output information to multiple destinations, such as the console, log files, or network dashboards. They often support configurable verbosity levels, allowing developers to filter logs based on importance, such as focusing only on warnings and errors in a production environment.

\section{Framework Design}
\label{sec:design}
Implementing a framework based on the ECS pattern requires an ECS-compatible architecture. This can be achieved in two different manners~\cite{buschmann2001pattern}. The first involves relying on a target ECS system. Frameworks built like this are specific to extending the capabilities of the target system. Consequently, they must be compatible with the system's original architectural styles and patterns.

The second involves creating or relying on a connection layer for integration with the infrastructure. This can be done in different ways: frameworks can implement a service-based layer to communicate with the base systems, for example (as long as it supports service-based coupling). Another way is to implement specific connectors for different system infrastructure aspects within the framework. This is the approach usually taken by game engines~\cite{gregory2017game}.

Game engines are frameworks or platforms designed to facilitate the development of video games. Due to their comprehensive nature, they can assume characteristics of both frameworks and platforms to the extent that it may be hard to evaluate whether a game engine is a framework or a platform. This happens due to the highly specialized nature of game engines: more than providing a set of libraries, they need complex systems for simulating real-world physics, processing in-game location, and handling multiple players—often in real-time.

As a result, it is common for game engines to implement infrastructure connectors related to networking and resource management. In this scenario, they retain their characteristics as frameworks (not using internal middleware or encapsulating the runtime environment) while possessing other system modules for efficient in-game messaging or texture rendering. This way, a game engine can embed the mechanisms necessary for online player-to-player communication while completely detached from the communication middleware.

Considering the complex nature of agent communication, we chose this approach to implement our framework. In order to use the ECS pattern while keeping it interoperable with different environments and middleware configurations, we included an \textit{engine layer} in our implementation. At this point, we were concerned about agent-to-agent communication using different middleware for specific interaction protocols (e.g., direct messaging or publish-subscriber communication~\cite{tanenbaum2007distributed}).

Another concern was related to different MAS implementations. We wanted our framework to be used in different scenarios without implementing specific mechanisms that would only be used in particular situations. 


With all these considerations in mind, we defined a set of guiding requirements to implement the proposed framework, listed below:

\begin{itemize}
    \item \textbf{GR1:} The framework must adhere to the ECS architectural pattern.
    \item \textbf{GR2:} The framework must provide complete support for agent-oriented abstractions 
    \item \textbf{GR3:} The framework should allow different instances of a MAS to communicate with each other independently of the underlying infrastructure. That means the framework, while using the ECS architectural pattern, must provide means for inter-system communication.
    \item \textbf{GR4:} Auxiliary external libraries must be kept to a bare minimum. This allows dependencies to specific MAS to be added at the system implementation level, outside the framework implementation.
\end{itemize}

These requirements were then broken into functional and non-functional requirements with the appropriate design constraints, as detailed in the next paragraphs.

\textbf{Functional Requirements} — 
\textbf{R1: Agent Abstraction Support} (different agent architectures, dynamic behavior specification, mental state management, autonomous decision-making); 
\textbf{R2: Organizational Structures} (dynamic group formation, role-based capabilities, organizational policies, multi-group membership); 
\textbf{R3: Communication Infrastructure} (message passing, ACL patterns, external broker integration, async/sync modes); 
\textbf{R4: Environmental Interaction} (perception/action execution, multiple environment models, spatial and non-spatial representations).

\textbf{Non-Functional Requirements} — 
\textbf{R5: Scalability} (distributed deployment, stateless agent processing, persistent state across restarts, scalable communication); 
\textbf{R6: Development} (config-based agent spec, Java implementation, IDE/tool integration, logging/debugging support); 
\textbf{R7: Integration} (message brokers like RabbitMQ/Kafka, RESTful APIs, JSON configuration, Docker support).

\textbf{Design Constraints} — 
\textbf{C1: Architectural Separation} (clear separation between agent logic and ECS details, communication infrastructure and reasoning, environment models and perceptions); 
\textbf{C2: Compatibility Requirements} (no reliance on game external engines, minimal external dependencies); 
\textbf{C3: Distribution Constraints} (stateless agent processing, pluggable persistent storage, communication resilience to network partitions, idempotent operations).

\subsection{Conceptual Mapping}
To satisfy the first two guiding requirements (GR1 and GR2), we need to preserve agent-oriented concepts while making them accessible through the familiar ECS pattern, eliminating the need for specialized agent languages or frameworks. This can be achieved through a \textit{conceptual mapping} between agent-oriented concepts and ECS elements, as shown in Table~\ref{tab:mapping}. 

\begin{table}[h]
\centering
\resizebox{\columnwidth}{!}{%
\begin{tabular}{@{}lll@{}}
\toprule
\textbf{Agent Concept} & \textbf{ECS Element} & \textbf{Implementation} \\
\midrule
Agent & Entity & Identifier with specific components \\
Agent properties & Components & Data containers (AgentComponent) \\
Agent behavior & Systems & Behavior systems and trees \\
Beliefs & Components & BeliefComponent with key-value store \\
Goals & Components & GoalComponent with priority queue \\
Communication & Components + Systems & MessageComponent + MessagingSystem \\
Organization & Components & GroupComponent with membership \\
Roles & Components & RoleComponent with capabilities \\
Environment & World + Systems & Environment representation and systems \\
\bottomrule
\end{tabular}%
}
\caption{Mapping between agent concepts and ECS elements}
\label{tab:mapping}
\end{table}

When reviewing existing work related to engineering MAS in the context of DS, we found a peer-reviewed set of abstractions that fit our needs perfectly. In~\cite{silva2002taming}, the authors establish a conceptual framework with distinct abstractions (and their relationships) to support the development of large-scale MASs. This framework, named TAO (Taming Agents and Objects), contains both consolidated abstractions (such as objects and classes) and “emergent” abstractions (such as agents, roles, and organizations).

The TAO framework emphasizes the separation of roles and agents, while modeling objects and teams as first-class concepts. Its key components can be described as follows:

\begin{itemize}
    \item Organizations: Represent collective structures where agents collaborate to achieve shared goals, supporting hierarchical or networked organizational arrangements to model complex multiagent systems. They also enable role-based organization: each agent within an organization has specific roles and responsibilities, which may vary depending on the team's objective or task.
    \item Agents: Represent autonomous entities capable of reasoning, acting, and interacting. Each agent has a decision-making mechanism that governs its actions, guided by its goals and perceptions. Agents can belong to multiple organizations, fostering flexibility in multi-organization collaboration.
    \item Objects: Represent environmental entities or resources with which agents interact. Objects also provide the context or operational space for the agents and organizations, and they may include both passive entities (e.g., tools, materials) and active systems (e.g., processes or services).
  \end{itemize}

The abstractions defined by TAO are classified as: 

\begin{itemize}
    \item Fundamental abstractions: include the object and agent abstractions ({\tt Agents} and {\tt Objects}).
    \item Environment abstractions: include the definition of environments and events that are used to represent its constraints and characteristics ({\tt Environment} and {\tt Event}).
    \item Grouping abstractions: abstractions for dealing with more complex situations in large-scale systems. It includes organizations and roles to model complex collaborations ({\tt Role} and {\tt Organization}).
\end{itemize}

Both ECS and TAO embrace the principle of separation of concerns: while ECS separates data ({\tt Components} and logic {\tt Systems}, TAO separates {\tt Agents}, {\tt Roles}, and {\tt Objects}, with {\tt Roles} being independent of the {\tt Agents} fulfilling them. They also both focus on collaboration: ECS systems can be designed for cooperative or collective behavior of entities, much like how TAO focuses on team-oriented agent collaboration. Since the TAO abstractions naturally align with the Entity-Component-System paradigm, we implemented a formal mapping between TAO abstractions and ECS elements. This mapping preserves agent-oriented semantics while leveraging ECS architectural benefits:

\begin{table}[h]
\centering
\begin{tabular}{@{}ll@{}}
\toprule
\textbf{TAO Abstraction} & \textbf{ECS Implementation} \\
\midrule
Object & Entity + ObjectComponent \\
Agent & Entity + AgentComponent + BeliefComponent \\
& + GoalComponent + IntentionComponent \\
Group & Entity + GroupComponent \\
Role & RoleComponent \\
\bottomrule
\end{tabular}
\caption{TAO to ECS mapping correspondences}
\label{tab:tao-mapping}
\end{table}

\section{Architecture}
\label{sec:architecture}

\subsection{Overall System Architecture}
\label{sec:overall-architecture}

HECATE employs a layered architecture that cleanly separates agent abstractions from the underlying ECS implementation. The architecture consists of three primary layers:

\textbf{Agent Abstraction Layer}: Provides high-level agent concepts and organizational structures. This layer implements TAO abstractions through specialized components and behaviors, maintaining semantic compatibility with agent theory while operating within ECS constraints.

\textbf{ECS Core Layer}: Implements the foundational Entity-Component-System pattern. This layer manages entity lifecycle, component storage, and system execution, optimized for performance and scalability.

\textbf{Infrastructure Layer}: Handles distributed systems concerns including communication, persistence, and monitoring. This layer integrates with external services such as message brokers and provides the foundation for horizontal scaling.

The architecture deliberately avoids tight coupling between layers, enabling independent modification and optimization of each component while maintaining overall system coherence.

\subsection{Core Component Design}
\label{sec:component-design}

HECATE implements agent concepts through specialized components that encapsulate agent state and properties:

\textbf{Entity Components} — 
\textbf{ObjectComponent} (base abstraction: \texttt{properties}, \texttt{objectType}, \texttt{behavior}); 
\textbf{AgentComponent} (extends ObjectComponent: \texttt{architecture}, \texttt{state}, \texttt{autonomyLevel}).

\textbf{Mental State Components} — 
\textbf{BeliefComponent} (beliefs: \texttt{beliefs}, \texttt{confidenceValues}, \texttt{revisionStrategy}); 
\textbf{GoalComponent} (goals: \texttt{goals}, \texttt{achievements}, \texttt{constraints}); 
\textbf{IntentionComponent} (intentions: \texttt{intentions}, \texttt{executionState}, \texttt{contingencies}).

\textbf{Organizational Components} — 
\textbf{GroupComponent} (organization: \texttt{members}, \texttt{policies}, \texttt{structure}); 
\textbf{RoleComponent} (roles: \texttt{roles}, \texttt{activeRole}, \texttt{permissions}).

\textbf{Agent Processing Systems} — 
\textbf{AgentSystem} (reasoning cycle, state transitions, BDI coordination); 
\textbf{PlanningSystem} (goal decomposition, plan libraries, dynamic planning).

\textbf{Organizational Systems} — 
\textbf{GroupSystem} (enforces group policies, intra-group communication, lifecycle management); 
\textbf{RoleSystem} (role assignment validation, permissions application, role conflict handling).

\textbf{Infrastructure Systems} — 
\textbf{MessagingSystem} (external broker integration, message routing, point-to-point/publish-subscribe patterns); 
\textbf{PersistenceSystem} (state snapshots, multi-backend storage, failure recovery).

\subsection{Communication Architecture}
\label{sec:communication-architecture}

HECATE's communication architecture leverages external message brokers to achieve scalability and reliability:

\subsubsection{Message Broker Integration}

The framework integrates with message brokers like RabbitMQ through a communication abstraction layer:
\begin{itemize}
    \item \textbf{Topic-based routing}: Messages are routed based on agent or group topics
    \item \textbf{Delivery guarantees}: Configurable delivery semantics (at-most-once, at-least-once)
    \item \textbf{Load balancing}: Automatic distribution of messages across agent instances
\end{itemize}

\subsubsection{Communication Patterns}

HECATE's communication architecture supports multiple message patterns through its integration with external message brokers. This flexibility enables different communication paradigms suitable for various agent interaction scenarios:
\begin{itemize}
    \item \textbf{Point-to-point}: Enables direct agent-to-agent communication with delivery guarantees.
    \item \textbf{Group multicasting}: Efficient message distribution to all group members.
    \item \textbf{Publish-subscribe}: Facilitates broadcasting to multiple agents through topic exchanges.
\end{itemize}

The external broker approach ensures that communication scalability is handled independently of agent processing, enabling horizontal scaling of both components.

\section{Implementation Overview}
\label{sec:implementation-overview}

HECATE implements a server-client architecture that fundamentally separates the agent processing infrastructure from the development environment. This separation enables the framework to achieve two key objectives: maintaining high-performance agent execution while providing accessible development interfaces across multiple programming languages.

\subsection{Server Architecture}
\label{sec:server-architecture}

The HECATE server constitutes the core processing engine, implemented in Java for optimal performance. The server's responsibilities include:

\begin{itemize}
    \item \textbf{Agent Lifecycle Management}: Creating, suspending, resuming, and terminating agents based on client requests
    \item \textbf{Reasoning Cycle Execution}: Processing agent behavior according to their architecture (reactive, cognitive, or BDI)
    \item \textbf{Environment Simulation}: Maintaining and updating the world state, handling agent perceptions and actions
    \item \textbf{Organizational Management}: Managing groups, roles, and relationships between agents
    \item \textbf{Message Brokering}: Coordinating agent communication through external message queues
    \item \textbf{State Persistence}: Maintaining agent states and enabling recovery from failures
\end{itemize}

The server operates as a stateful service that maintains the complete agent world model. It exposes RESTful APIs for agent management and WebSocket connections for real-time message delivery. All computationally intensive operations, including agent reasoning cycles and environment updates, occur within the server's ECS engine.

\subsection{SDK Architecture}
\label{sec:sdk-architecture}

The SDK provides lightweight client libraries that facilitate agent development and interaction. Currently available in Java with planned extensions to other languages, the SDK's responsibilities include:

\begin{itemize}
    \item \textbf{Agent Specification}: Providing domain-specific language or configuration interfaces for defining agent behaviors
    \item \textbf{Server Communication}: Managing connections to the HECATE server through HTTP and WebSocket protocols
    \item \textbf{Event Handling}: Translating server events into native language constructs
    \item \textbf{Behavior Definition}: Offering abstractions for defining agent behaviors without ECS implementation details
    \item \textbf{Tool Integration}: Enabling integration with standard development environments and debugging tools
\end{itemize}

The SDK acts as a facade, hiding the complexities of the underlying ECS architecture while exposing agent-oriented programming constructs. Developers work with familiar object-oriented or functional paradigms within their chosen language, with the SDK translating these abstractions into server-compatible formats.

\textbf{Server–SDK Communication} — 
\textbf{REST API} (agent creation/configuration, group and role management, agent state queries, environment interaction requests); 
\textbf{WebSocket Connection} (real-time messaging, perception updates, async event notifications, bidirectional behavior communication).





This dual-protocol approach optimizes for both management operations (REST) and real-time interaction (WebSockets), ensuring efficient resource utilization while maintaining responsive agent behavior.


\subsection{Deployment}
\label{sec:deployment}

HECATE's server-client architecture enables flexible deployment configurations suitable for various scales and requirements. The framework separates agent processing (server) from agent development (SDK), allowing developers to implement agents in their preferred programming languages while benefiting from centralized infrastructure. The HECATE server can be deployed as a standalone application or containerized using Docker, while multi-server clusters are supported through message broker federation. MAS applications connect to the server through lightweight SDK libraries. 

\section{Case Study}
\label{sec:case-study}

We evaluated HECATE as an educational tool in two graduate-level courses:

\begin{itemize}
    \item \textbf{Course 1}: Used HECATE as a framework for implementing MAS concepts
    \item \textbf{Course 2}: Focused on teaching MAS engineering using distributed systems principles
\end{itemize}

Both courses targeted students with no prior knowledge of agent theory but with varying levels of programming experience. Students were tasked with implementing multi-agent systems to solve specific problems, using either the HECATE framework (Course 1) or distributed systems principles applied to agent concepts (Course 2).

The results of this study were discussed and published in~\cite{dummyRef} (omitted for review). In summary, we concluded that our approach successfully leveraged the learning curve for MAS education by incorporating agent concepts into familiar distributed systems patterns.

\section{Discussion}
\label{sec:discussion}

\subsection{Benefits of the ECS Approach}
The HECATE framework offers several advantages for MAS development and education:

\begin{itemize}
    \item \textbf{Accessibility}: By mapping agent concepts to the familiar ECS pattern, HECATE reduces the learning curve for developers and students
    \item \textbf{Integration}: The framework integrates seamlessly with existing distributed systems practices and tools
    \item \textbf{Scalability}: The data-oriented design approach enables efficient handling of large numbers of agents
    \item \textbf{Flexibility}: Different agent architectures (reactive, cognitive, BDI) can be implemented through component composition
    \item \textbf{Simplicity}: Declarative agent definition through JSON reduces complexity
\end{itemize}

\subsection{Comparison with Traditional MAS Frameworks}
HECATE differs from traditional MAS frameworks in several key aspects:

\begin{table}[h]
\centering
\resizebox{\columnwidth}{!}{%
\begin{tabular}{@{}lll@{}}
\toprule
\textbf{Aspect} & \textbf{Traditional MAS Frameworks} & \textbf{HECATE} \\
\midrule
Programming paradigm & Agent-oriented & Data-oriented \\
Learning curve & Steep & Moderate \\
Integration with DS & Limited & Native \\
Agent definition & Specialized languages & JSON configuration \\
Extensibility & Framework-specific & Component-based \\
Performance & Variable & Optimized for data locality \\
\bottomrule
\end{tabular}%
}
\caption{Comparison between traditional MAS frameworks and HECATE}
\label{tab:comparison}
\end{table}

While traditional frameworks provide deeper agent-specific features, HECATE prioritizes accessibility and integration with mainstream software engineering practices. The architectural decision to separate server infrastructure from client SDKs provides significant pedagogical and practical advantages. Engineers can work in their preferred programming languages without sacrificing agent-oriented concepts by abstracting the core agent processing in the server while providing lightweight SDKs. Currently, HECATE provides a comprehensive Java SDK that demonstrates full framework capabilities, including reactive, cognitive, and BDI agent architectures. SDKs for Python, JavaScript, and Kotlin are planned for upcoming releases, allowing broader adoption across different educational contexts and programming backgrounds. This approach eliminates the language barrier that often impedes the teaching of multi-agent systems, while maintaining the theoretical rigor of agent-oriented programming. The server-client separation also facilitates hybrid agent systems where different components can be implemented in the most appropriate languages for specific tasks.

\subsection{Limitations}
From the MAS perspective, HECATE still has some limitations that should be addressed in future work:

\begin{itemize}
    \item \textbf{Advanced Agent Features}: Some specialized MAS features (e.g., complex norm reasoning) are not yet fully implemented
    \item \textbf{Formal Verification}: Unlike some agent-specific languages, HECATE does not support formal verification of agent properties
    \item \textbf{Standardization}: The framework does not yet comply with agent communication standards like FIPA-ACL
    \item \textbf{Distributed Operation}: While designed for distribution, the current implementation focuses on single-node operation
\end{itemize}

\subsection{Educational Implications}
Our case studies demonstrate that HECATE can significantly improve the teaching of MAS concepts by:

\begin{itemize}
    \item Providing a concrete, familiar architectural pattern for implementing agent concepts
    \item Reducing the cognitive load associated with learning specialized agent languages
    \item Connecting agent theory to mainstream software engineering practices
    \item Allowing incremental introduction of advanced agent concepts
\end{itemize}

This approach addresses the challenge identified in~\cite{dummyRef} of making MAS more accessible to the broader software engineering community.

\subsection{Conclusion and Future Work}
In this paper, we presented HECATE, an ECS-based framework for teaching and developing multi-agent systems that bridges the gap between distributed systems engineering and MAS development. By mapping agent concepts to the ECS architectural pattern, HECATE makes MAS development more accessible to students and practitioners familiar with mainstream software engineering practices.

Implementing the messaging module posed challenges in maintaining the decoupled nature of ECS. For this reason, it was implemented as a separate layer to ensure that communication between agents remained outside the core ECS architecture. One advantage of taking this approach is related to incorporating different messaging mechanisms when necessary: considering our DS-based approach, there may be situations where different mechanisms are necessary to accommodate both inter- and intra-agent communication. Separating the entire mechanism within its own layer ensures that adapting different messaging mechanisms can be done without impacting the rest of the framework.

In terms of practical applications, our case study demonstrated that HECATE successfully reduces the learning curve for MAS development while maintaining the core concepts necessary for agent-oriented programming. Students without prior agent knowledge achieved high proficiency in MAS concepts in less time compared to traditional approaches.

Future work on HECATE will focus on:

\begin{itemize}
    \item Expanding the framework to support distributed operation across multiple nodes
    \item Implementing advanced agent features such as argumentation and normative reasoning
    \item Adding support for standard agent communication languages
    \item Developing visual tools for agent design and monitoring
    \item Conducting larger-scale evaluations in educational and industrial settings
\end{itemize}

HECATE demonstrates that incorporating MAS concepts into distributed systems architectural patterns can significantly lower the barrier to entry for MAS development and education, potentially increasing the adoption of agent-oriented approaches in mainstream software engineering.



\begin{ack}
This work was supported in part by [funding information].
\end{ack}



\bibliography{ecai}

\begin{thebibliography}{52}
\providecommand{\natexlab}[1]{#1}
\providecommand{\url}[1]{\texttt{#1}}
\expandafter\ifx\csname urlstyle\endcsname\relax
  \providecommand{\doi}[1]{doi: #1}\else
  \providecommand{\doi}{doi: \begingroup \urlstyle{rm}\Url}\fi

\bibitem[Ahmad(2002)]{ahmad2002multi}
H.~F. Ahmad.
\newblock Multi-agent systems: overview of a new paradigm for distributed systems.
\newblock In \emph{7th IEEE International Symposium on High Assurance Systems Engineering, 2002. Proceedings.}, pages 101--107. IEEE, 2002.

\bibitem[Anonymous(XXXX)]{dummyRef}
Anonymous.
\newblock Title omitted for blind review.
\newblock \emph{Journal omitted for blind review}, XX\penalty0 (X):\penalty0 XX--XX, XXXX.

\bibitem[Baldoni et~al.(2021)Baldoni, Baroglio, Micalizio, and Tedeschi]{baldoni2021reimagining}
M.~Baldoni, C.~Baroglio, R.~Micalizio, and S.~Tedeschi.
\newblock Reimagining robust distributed systems through accountable mas.
\newblock \emph{IEEE Internet Computing}, 25\penalty0 (6):\penalty0 7--14, 2021.

\bibitem[Baron(2019)]{baron2019hands}
D.~Baron.
\newblock \emph{Hands-On Game Development Patterns with Unity 2019}.
\newblock Packt Publishing, 2019.

\bibitem[Bellifemine et~al.(2007)Bellifemine, Caire, and Greenwood]{bellifemine2007developing}
F.~L. Bellifemine, G.~Caire, and D.~Greenwood.
\newblock \emph{Developing multi-agent systems with JADE}, volume~7.
\newblock John Wiley \& Sons, 2007.

\bibitem[Boissier et~al.(2013)Boissier, Bordini, H{\"u}bner, Ricci, and Santi]{boissier2013multi}
O.~Boissier, R.~H. Bordini, J.~F. H{\"u}bner, A.~Ricci, and A.~Santi.
\newblock Multi-agent oriented programming with jacamo.
\newblock \emph{Science of Computer Programming}, 78\penalty0 (6):\penalty0 747--761, 2013.

\bibitem[Briola et~al.(2023)Briola, Ferrando, and Mascardi]{mascardifantastic}
D.~Briola, A.~Ferrando, and V.~Mascardi.
\newblock Fantastic mass and where to find them: First results and lesson learned.
\newblock In \emph{International Workshop on Engineering Multi-Agent Systems}, pages 233--252. Springer, 2023.

\bibitem[Burns and Oppenheimer(2016)]{burns2016design}
B.~Burns and D.~Oppenheimer.
\newblock Design patterns for container-based distributed systems.
\newblock In \emph{8th USENIX Workshop on Hot Topics in Cloud Computing (HotCloud 16)}, 2016.

\bibitem[Buschmann et~al.(2001)Buschmann, Meunier, Rohnert, Sornmerlad, and Stal]{buschmann2001pattern}
F.~Buschmann, R.~Meunier, H.~Rohnert, P.~Sornmerlad, and M.~Stal.
\newblock \emph{Pattern-oriented software architecture: a system of patterns. Volume 1}.
\newblock Wiley, 2001.

\bibitem[Cabri et~al.(2007)Cabri, Leonardi, and Puviani]{cabri2007service}
G.~Cabri, L.~Leonardi, and M.~Puviani.
\newblock Service-oriented agent methodologies.
\newblock In \emph{Enabling Technologies: Infrastructure for Collaborative Enterprises, 2007. WETICE 2007. 16th IEEE International Workshops on}, pages 24--29. IEEE, 2007.

\bibitem[Chopra(2022)]{chopra2022interaction}
A.~K. Chopra.
\newblock Interaction-oriented software engineering: Programming abstractions for autonomy and decentralization.
\newblock \emph{AI Communications}, 35\penalty0 (4):\penalty0 381--391, 2022.

\bibitem[Chopra et~al.(2021{\natexlab{a}})Chopra, Christie~V, and Singh]{chopra2021multiagent}
A.~K. Chopra, S.~H. Christie~V, and M.~P. Singh.
\newblock Multiagent foundations for distributed systems: A vision.
\newblock In \emph{International Workshop on Engineering Multi-Agent Systems}, pages 62--71. Springer, 2021{\natexlab{a}}.

\bibitem[Chopra et~al.(2021{\natexlab{b}})Chopra, Singh, et~al.]{chopra2021deserv}
A.~K. Chopra, M.~P. Singh, et~al.
\newblock Deserv: Decentralized serverless computing.
\newblock In \emph{2021 IEEE International Conference on Web Services (ICWS)}, pages 51--60. IEEE, 2021{\natexlab{b}}.

\bibitem[Christie et~al.(2022)Christie, Chopra, and Singh]{christie2022mandrake}
S.~H. Christie, A.~K. Chopra, and M.~P. Singh.
\newblock Mandrake: multiagent systems as a basis for programming fault-tolerant decentralized applications.
\newblock \emph{Autonomous Agents and Multi-Agent Systems}, 36\penalty0 (1):\penalty0 16, 2022.

\bibitem[Christie et~al.(2023)Christie, Singh, and Chopra]{christie2023kiko}
S.~H. Christie, M.~P. Singh, and A.~K. Chopra.
\newblock Kiko: programming agents to enact interaction protocols.
\newblock In \emph{Proceedings of the International Conference on Autonomous Agents and MultiAgent Systems (AAMAS}, volume~22, 2023.

\bibitem[Ciortea et~al.(2017{\natexlab{a}})Ciortea, Boissier, and Ricci]{ciortea2017beyond}
A.~Ciortea, O.~Boissier, and A.~Ricci.
\newblock Beyond physical mashups: Autonomous systems for the web of things.
\newblock In \emph{Proceedings of the Eighth International Workshop on the Web of Things}, pages 16--20. ACM, 2017{\natexlab{a}}.

\bibitem[Ciortea et~al.(2017{\natexlab{b}})Ciortea, Boissier, Zimmermann, and Florea]{Ciortea:2017:GAR:3091125.3091342}
A.~Ciortea, O.~Boissier, A.~Zimmermann, and A.~M. Florea.
\newblock Give agents some rest: A resource-oriented abstraction layer for internet-scale agent environments.
\newblock In \emph{Proceedings of the 16th Conference on Autonomous Agents and MultiAgent Systems}, AAMAS '17, pages 1502--1504, 2017{\natexlab{b}}.

\bibitem[Collier et~al.(2015)Collier, Russell, and Lillis]{collier2015reflecting}
R.~W. Collier, S.~Russell, and D.~Lillis.
\newblock Reflecting on agent programming with agentspeak (l).
\newblock In \emph{PRIMA 2015: Principles and Practice of Multi-Agent Systems: 18th International Conference, Bertinoro, Italy, October 26-30, 2015, Proceedings 13}, pages 351--366. Springer, 2015.

\bibitem[Crnkovic and Larsson(2002)]{crnkovic2002building}
I.~Crnkovic and M.~P.~H. Larsson.
\newblock \emph{Building reliable component-based software systems}.
\newblock Artech House, 2002.

\bibitem[Dzitac and B{\u{a}}rbat(2009)]{dzitac2009artificial}
I.~Dzitac and B.~E. B{\u{a}}rbat.
\newblock Artificial intelligence+ distributed systems= agents.
\newblock \emph{International Journal of Computers Communications \& Control}, 4\penalty0 (1):\penalty0 17--26, 2009.

\bibitem[Fabian(2020)]{fabian2020data}
R.~Fabian.
\newblock \emph{Data-Oriented Design}.
\newblock Self-published, 2020.

\bibitem[Fedoseev et~al.(2020)Fedoseev, Askarbekuly, Uzbekova, and Mazzara]{fedoseev2020case}
K.~Fedoseev, N.~Askarbekuly, E.~Uzbekova, and M.~Mazzara.
\newblock A case study on object-oriented and data-oriented design paradigms in game development.
\newblock 2020.

\bibitem[Ferber(1999)]{ferber1999multi}
J.~Ferber.
\newblock \emph{Multi-agent systems: an introduction to distributed artificial intelligence}, volume~1.
\newblock Addison-Wesley Reading, 1999.

\bibitem[Fortino et~al.(2014)Fortino, Guerrieri, Russo, and Savaglio]{fortino2014integration}
G.~Fortino, A.~Guerrieri, W.~Russo, and C.~Savaglio.
\newblock Integration of agent-based and cloud computing for the smart objects-oriented iot.
\newblock In \emph{Proceedings of the 2014 IEEE 18th international conference on computer supported cooperative work in design (CSCWD)}, pages 493--498. IEEE, 2014.

\bibitem[Gregory(2017)]{gregory2017game}
J.~Gregory.
\newblock \emph{Game Engine Architecture}.
\newblock CRC Press, 3rd edition, 2017.

\bibitem[Hopgood(2021)]{hopgood2021intelligent}
A.~A. Hopgood.
\newblock \emph{Intelligent systems for engineers and scientists: a practical guide to artificial intelligence}.
\newblock CRC press, 2021.

\bibitem[Khadse et~al.(2023)Khadse, Christie~V, Singh, and Chopra]{khadse2023protocol}
A.~K. Khadse, S.~H. Christie~V, M.~P. Singh, and A.~K. Chopra.
\newblock Protocol-based engineering of microservices.
\newblock In \emph{International Workshop on Engineering Multi-Agent Systems}, pages 61--77. Springer, 2023.

\bibitem[Khosla and Ichalkaranje(2013)]{khosla2013design}
R.~Khosla and N.~Ichalkaranje.
\newblock \emph{Design of intelligent multi-agent systems: human-centredness, architectures, learning and adaptation}.
\newblock Springer, 2013.

\bibitem[Lange(1998)]{lange1998mobile}
D.~B. Lange.
\newblock Mobile objects and mobile agents: The future of distributed computing?
\newblock In \emph{European conference on object-oriented programming}, pages 1--12. Springer, 1998.

\bibitem[Leon et~al.(2015)Leon, Paprzycki, and Ganzha]{leon2015review}
F.~Leon, M.~Paprzycki, and M.~Ganzha.
\newblock A review of agent platforms.
\newblock \emph{Multi-paradigm Modelling for Cyber-Physical Systems (MPM4CPS), ICT COST Action IC1404}, pages 1--15, 2015.

\bibitem[Mascardi et~al.(2019)Mascardi, Weyns, and Ricci]{Mascardi2019}
V.~Mascardi, D.~Weyns, and A.~Ricci.
\newblock Engineering multi-agent systems: State of affairs and the road ahead.
\newblock \emph{ACM SIGSOFT Software Engineering Notes}, 44\penalty0 (1):\penalty0 18--28, 2019.

\bibitem[McGrath and Brenner(2017)]{mcgrath2017serverless}
G.~McGrath and P.~R. Brenner.
\newblock Serverless computing: Design, implementation, and performance.
\newblock In \emph{2017 IEEE 37th International Conference on Distributed Computing Systems Workshops (ICDCSW)}, pages 405--410. IEEE, 2017.

\bibitem[Meyer(1997)]{meyer1997object}
B.~Meyer.
\newblock \emph{Object-oriented software construction}, volume~2.
\newblock Prentice hall Englewood Cliffs, 1997.

\bibitem[Mironov et~al.(2021)Mironov, Motaylenko, Andreev, Antonov, and Aristov]{mironov2021comparison}
T.~Mironov, L.~Motaylenko, D.~Andreev, I.~Antonov, and M.~Aristov.
\newblock Comparison of object-oriented programming and data-oriented design for implementing trading strategies backtester.
\newblock In \emph{ENVIRONMENT. TECHNOLOGIES. RESOURCES. Proceedings of the International Scientific and Practical Conference}, volume~2, pages 124--130, 2021.

\bibitem[Newman(2015)]{newman2015building}
S.~Newman.
\newblock \emph{Building microservices: designing fine-grained systems}.
\newblock " O'Reilly Media, Inc.", 2015.

\bibitem[Nissim and Brafman(2012)]{nissim2012multi}
R.~Nissim and R.~I. Brafman.
\newblock Multi-agent a* for parallel and distributed systems.
\newblock In \emph{AAMAS}, pages 1265--1266, 2012.

\bibitem[Nystrom(2014)]{nystrom2014game}
R.~Nystrom.
\newblock \emph{Game Programming Patterns}.
\newblock Genever Benning, 2014.

\bibitem[O’Neill et~al.(2020)O’Neill, Lillis, O’Hare, and Collier]{o2020delivering}
E.~O’Neill, D.~Lillis, G.~M. O’Hare, and R.~W. Collier.
\newblock Delivering multi-agent microservices using cartago.
\newblock In \emph{Engineering Multi-Agent Systems: 8th International Workshop, EMAS 2020, Auckland, New Zealand, May 8--9, 2020, Revised Selected Papers 8}, pages 1--20. Springer, 2020.

\bibitem[Palanca et~al.(2020)Palanca, Terrasa, Julian, and Carrascosa]{palanca2020spade}
J.~Palanca, A.~Terrasa, V.~Julian, and C.~Carrascosa.
\newblock Spade 3: Supporting the new generation of multi-agent systems.
\newblock \emph{IEEE Access}, 8:\penalty0 182537--182549, 2020.

\bibitem[Pfeifer et~al.(2021)Pfeifer, Passini, Dorante, Guilherme, and Affonso]{pfeifer2021multi}
V.~Pfeifer, W.~F. Passini, W.~F. Dorante, I.~R. Guilherme, and F.~J. Affonso.
\newblock A multi-agent approach to monitor and manage container-based distributed systems.
\newblock \emph{IEEE Latin America Transactions}, 20\penalty0 (1):\penalty0 82--91, 2021.

\bibitem[Rao et~al.(1995)Rao, Georgeff, et~al.]{rao1995bdi}
A.~S. Rao, M.~P. Georgeff, et~al.
\newblock Bdi agents: from theory to practice.
\newblock In \emph{ICMAS}, volume~95, pages 312--319, 1995.

\bibitem[Rose et~al.(2015)Rose, Eldridge, and Chapin]{rose2015internet}
K.~Rose, S.~Eldridge, and L.~Chapin.
\newblock The internet of things: An overview.
\newblock \emph{The internet society (ISOC)}, 80\penalty0 (15):\penalty0 1--53, 2015.

\bibitem[Russell and Norvig(1995)]{Russell:1995:AIM:193191}
S.~J. Russell and P.~Norvig.
\newblock \emph{Artificial Intelligence: A Modern Approach}.
\newblock Prentice-Hall, Inc., Upper Saddle River, NJ, USA, 1995.
\newblock ISBN 0-13-103805-2.

\bibitem[Salah et~al.(2016)Salah, Zemerly, Yeun, Al-Qutayri, and Al-Hammadi]{salah2016evolution}
T.~Salah, M.~J. Zemerly, C.~Y. Yeun, M.~Al-Qutayri, and Y.~Al-Hammadi.
\newblock The evolution of distributed systems towards microservices architecture.
\newblock In \emph{2016 11th International Conference for Internet Technology and Secured Transactions (ICITST)}, pages 318--325. IEEE, 2016.

\bibitem[Silva et~al.(2002)Silva, Garcia, Brand{\~a}o, Chavez, Lucena, and Alencar]{silva2002taming}
V.~Silva, A.~Garcia, A.~Brand{\~a}o, C.~Chavez, C.~Lucena, and P.~Alencar.
\newblock Taming agents and objects in software engineering.
\newblock In \emph{International Workshop on Software Engineering for Large-Scale Multi-agent Systems}, pages 1--26. Springer, 2002.

\bibitem[Su and Wu(2011)]{su2011jade}
C.-J. Su and C.-Y. Wu.
\newblock Jade implemented mobile multi-agent based, distributed information platform for pervasive health care monitoring.
\newblock \emph{Applied Soft Computing}, 11\penalty0 (1):\penalty0 315--325, 2011.

\bibitem[Tanenbaum and van Steen(2007)]{tanenbaum2007distributed}
A.~S. Tanenbaum and M.~van Steen.
\newblock \emph{Distributed Systems: Principles and Paradigms}.
\newblock Prentice-Hall, 2007.

\bibitem[Van~Steen and Tanenbaum(2017)]{steen2017distributed}
M.~Van~Steen and A.~S. Tanenbaum.
\newblock \emph{Distributed Systems}.
\newblock distributed-systems.net, 3 edition, 2017.
\newblock ISBN 1543057381.

\bibitem[W.~Collier et~al.(2019)W.~Collier, O'Neill, Lillis, and O'Hare]{w2019mams}
R.~W.~Collier, E.~O'Neill, D.~Lillis, and G.~O'Hare.
\newblock Mams: Multi-agent microservices.
\newblock In \emph{Companion proceedings of the 2019 world wide web conference}, pages 655--662, 2019.

\bibitem[Weiss(2016)]{weiss2016multiagent}
G.~Weiss.
\newblock \emph{Multiagent Systems: A Modern Approach to Distributed Artificial Intelligence}.
\newblock MIT Press, 2016.

\bibitem[Weyns(2010)]{weyns2010architecture}
D.~Weyns.
\newblock \emph{Architecture-based design of multi-agent systems}.
\newblock Springer Science \& Business Media, 2010.

\bibitem[Wiebusch and Latoschik(2015)]{wiebusch2015decoupling}
D.~Wiebusch and M.~E. Latoschik.
\newblock Decoupling the entity-component-system pattern using semantic traits for reusable realtime interactive systems.
\newblock In \emph{2015 IEEE 8th Workshop on Software Engineering and Architectures for Realtime Interactive Systems (SEARIS)}, pages 25--32. IEEE, 2015.

\end{thebibliography}

\end{document}